\documentclass[runningheads,a4paper]{llncs}
\usepackage{graphicx}
\usepackage[cmex10]{amsmath}
\usepackage{amssymb}
\usepackage{array}
\usepackage{url}
\usepackage{subfig}
\usepackage{hyphenat}
\renewcommand{\figurename}{Figure}

\include{Qcircuit}

\begin{document}

\title{A Fully Fault-Tolerant Representation of Quantum Circuits}

\author{Alexandru Paler$^*$\and Ilia Polian$^*$\and Kae Nemoto$^{**}$\and Simon J. Devitt$^{+,-,**}$}
%
\authorrunning{A Fully Fault-Tolerant Representation of Quantum Circuits}

\institute{$^*$University of Passau, Innstr. 43, 94032, Passau, Germany\\
$^+$Ochanomizu University, 2-1-1 Otsuka, Bunkyo-ku, Tokyo 112-8610, Japan. \\
$^-$Graduate School of Media and Governance, Keio University, Fujisawa, Kanagawa 252-0882, Japan.\\
$^{**}$National Institute of Informatics, 2-1-2 Hitotsubashi, Chiyoda-ku, Tokyo, Japan
}

\maketitle

\begin{abstract}
We present a quantum circuit representation consisting entirely of qubit initialisations (I), a network of controlled-NOT gates (C) and measurements with respect to different bases (M). The ICM representation is useful for optimisation of quantum circuits that include teleportation, which is required for fault-tolerant, error corrected quantum computation. The non-deterministic nature of teleportation necessitates the conditional introduction of corrective quantum gates and additional ancillae during circuit execution. Therefore, the standard optimisation objectives, gate count and number of wires, are not well-defined for general teleportation-based circuits. The transformation of a circuit into the ICM representation provides a canonical form for an exact fault-tolerant, error corrected circuit needed for optimisation prior to the final implementation in a realistic hardware model.
\end{abstract}

\section{Introduction}
\label{sec:intro}

Quantum computing promises speed-ups for a number of relevant computational problems. Building a scalable and reliable quantum computer is one of the challenges of modern science. As the size of quantum computers increases, the focus of interest shifts from their basic physical principles to structured design methodologies that will allow us to realise large-scale systems.

In general, quantum circuit optimisation methods are used to minimise the implementation costs like the number of gates or the number of wires~\cite{wille2010towards}. Classical circuit optimisation assumes fixed gate lists even in the presence of gate errors, but classical circuits are more robust towards errors, whereas quantum information is fragile~\cite[Ch.~8]{NC00}. Classical gate failures are usually solved either by hardening the circuit (e.g. modifying transistor sizes), or by introducing various types of information redundancies that mitigate the failures. Gate hardening is not considered realistic in quantum computing architectures, and a feasible solution requires quantum error-correcting codes (QECC)~\cite{devitt2013quantum}. The structure and design 
of QECC allows encoded quantum gates to be applied directly to the encoded quantum data. 

In contrast to the classical case, the most practical implementations of QECC and fault-tolerant quantum circuits are composed of gates which are \emph{non\hyp{}deterministic} even in the absence of errors~\cite{FMM13}. They either work correctly or require a correction, which is only determined during the execution of the circuit. Most such correction gates do not need to be dynamically included into the executing circuit, because their effect can be classically tracked through the subsequent gates~\cite{paler2014software}.  This is not true for all possible corrections occurring during the 
execution of a quantum circuit and some need to be actively applied to the quantum data\cite{FMM13}.  This means that the overall circuit is dynamic, because its gate list needs to be modified during its execution based on certain measurement results. Reducing the incidence of such gates is difficult because when a fully error-corrected, fault-tolerant circuit is examined, it is exactly these measurement based corrections that appear to give 
quantum computing its power~\cite{fowler2012time}. In general, fault-tolerant quantum circuits are constructed from Clifford and $T$ (Section~\ref{sec:quant}) gates, and the $T$ gate is the main source of the complications~\cite{amy2013meet} for which dynamic corrections cannot be avoided.

The separation of circuit gates into Clifford and $T$ gates is generally performed at the higher level circuit design layer in order to make 
fault-tolerant error constructions more amenable to practical implementation.  The physical mapping of these circuits to an actual error 
corrected architecture is then done with a specific QECC and hardware architecture in mind, preserving fault-tolerance. Fault-tolerance is understood as the set of procedures by which the cascade of quantum errors (bit and phase flips) caused by the 
circuit~\cite{devitt2013quantum} is restricted allowing the 
underlying QECC to be effective when mapped to actual operations in a hardware model.  In standard fault-tolerant constructions 
(those that are widely used in state-of-the-art hardware models~\cite{devitt2009architectural,Y13,N14,J13}), the only dynamic corrections 
needed are when we implement 
logical layer corrections for $T$ gates. These correctional gates are constructed using ancillae initialised into high-fidelity states 
(see Section~\ref{sec:igtele}) and gate teleportation protocols~\cite{FMM13}.  Our results are quite similar to those present in Ref. 
\cite{danos2007measurement}, however this work focuses on producing a representation that is compatible with fault-tolerant 
error correction protocols.

The solution to having all the required corrections into the logical layer of the computation is to translate circuits into a regular representation that replaces correctional gate dynamics with the dynamics of reading and interpreting the circuit outputs. Such an approach is similar to the model of measurement based quantum computing (MBQC)~\cite{briegel2009measurement}, where a computation is solely described by the interpretation of the measurements performed on a specifically initialised quantum state. A circuit is described in this work as an $ICM$ sequence, where the $I$ part contains qubit initialisations, the $C$ part is a sub-circuit consisting entirely of CNOT gates, and the qubits are measured in the $M$ part. This work represents a separate and distinct approach from the work of~\cite{miller2014mapping}, where $NCV$ (reversible) circuits were mapped into Clifford and $T$ gate circuits, because the ICM representation is regular and consists entirely of ancillae, CNOTs and measurements.

The ICM representation is the extension of the methods presented in~\cite{fowler2012time} to fit into the measurement based paradigm~\cite{briegel2009measurement}. The presented algorithmic formulation will output the ICM representation for arbitrary quantum and reversible circuits. Such a formulation, although it requires an increased number of ancillae, allows us to directly synthesise fully fault-tolerant 
error corrected circuits for an underlying higher level circuit (including all required ancillary protocols), represents the realistic resource requirements of fault-tolerant quantum computations for state-of-the-art quantum architectures~\cite{devitt2013requirements,gottesman2013overhead} and provides an elegant form for 
further circuit optimisation techniques for QECC models such as topological codes~\cite{PF13,mequanics}. 

The paper is organised as follows: Section~\ref{sec:quant} offers a short introduction to quantum computing, illustrates the concepts of controlled and rotational gates, discusses the reversibility aspects of computing and the applications of information and gate teleportations. Section~\ref{sec:materials} details the non-deterministic resource requirements of arbitrary quantum circuits, introduces the ICM representation and presents the algorithm used for achieving it. The algorithm is benchmarked using circuits from the RevLib library and the results are discussed in Section~\ref{sec:discussion}. Finally, conclusions and future work are formulated.

\subsection{Quantum and reversible computing}
\label{sec:quant}

\emph{Quantum circuits} represent and manipulate information in \emph{qubits} (quantum bits). The \emph{quantum state} of a qubit is the vector $\ket{\psi}= (\alpha_0, \alpha_1)^T = \alpha_0\ket{0} + \alpha_1\ket{1}$. Here, $\ket0 = (1, 0)^T$ and $\ket1 = (0, 1)^T$ are quantum analogues of classical logic values 0 and 1, respectively. $\alpha_0$ and $\alpha_1$ are complex numbers called \emph{amplitudes} with $|\alpha_0|^2 +|\alpha_1|^2 = 1$.

A state may be modified by applying single-qubit \emph{quantum gates}. Each quantum gate corresponds to a complex unitary matrix, and gate function is given by multiplying that matrix with the quantum state. The application of $X$ gate to a state results in a \emph{bit flip}: $X(\alpha_0, \alpha_1)^T = (\alpha_1, \alpha_0)^T$. The application of the $Z$ gate results in a \emph{phase flip}: $Z(\alpha_0, \alpha_1)^T= (\alpha_0, -\alpha_1)^T$. The matrices of the Pauli gates $I,X,Y,Z$ are:
\begin{eqnarray*}
\small{
\begin{array}{cccccccccccc}
I & = &
\left( \begin{array}{cc}
	1 & 0\\
	0 & 1
	\end{array} \right)
&
 Y & = &
\left( \begin{array}{cc}
	0 & -i\\
	i & 0
	\end{array} \right)
&
X & = &
\left( \begin{array}{cc}
	0 & 1\\
	1 & 0
	\end{array} \right)
&
 Z & = &
\left( \begin{array}{cc}
	1 & 0\\
	0 & -1
	\end{array} \right)
\end{array}}
\end{eqnarray*}

Further important single-qubit quantum gates in the context of this work are $H,P,T$, where $T^2 = P$ and $P^2 = Z$.
\begin{eqnarray*}
\small
H = \frac1{\sqrt2}\begin{pmatrix}
1 & 1 \\
1 & -1
\end{pmatrix}\,\,
P = \begin{pmatrix}
1 & 0 \\
0 & i
\end{pmatrix} \,\,
T = \begin{pmatrix}
1 & 0 \\
0 & e^{i\frac{\pi}{4}}
\end{pmatrix}
\end{eqnarray*}

\emph{Quantum measurement} is defined with respect to a basis and yields one of the basis vectors with a probability related to the amplitudes of the quantum state. Of importance in this work are $Z$- and $X$-measurements. $Z$-measurement is defined with respect to basis $(\ket0, \ket1)$. Applying a $Z$-measurement to a qubit in state $\ket{\psi}= \alpha_0\ket{0} + \alpha_1\ket{1}$ yields $\ket0$ with probability $|\alpha_0|^2$ and $\ket1$ with probability $|\alpha_1|^2$. Moreover, the state $\ket\psi$ \emph{collapses} into the measured state (i.e. only the components 
of $\ket{\psi}$ consistent with the measurement result remains). $X$-measurement is defined with respect to the basis $(\ket+, \ket-)$, where $\ket{+} = \frac{1}{\sqrt2}(\ket{0} + \ket{1})$ and $\ket{-}=\frac{1}{\sqrt2}(\ket{0}-\ket{1})$.

\subsection{Rotational gates}
The exponentiation of the Pauli matrices results in the rotational gates $R_x$, $R_y$, $R_z$ parametrised by the angle of the rotation~\cite[Ch.~4]{NC00}. Hence the bit flip is a rotation by $\pi$ around the $X$-axis, implying that $X=R_x(\pi)$, and the phase-flip is a rotation by $\pi$ around the $Z$-axis, such that $Z=R_z(\pi)$. Furthermore, $P=R_z(\pi/2)$ and $T=R_z(\pi/4)$. The $V$ and $V^\dagger$ gates are parametrised $X$-rotations, $V=R_x(\pi/2)$. The Hadamard gate is $H=R_z(\pi/2)R_x(\pi/2)R_z(\pi/2)=PVP$.

\begin{eqnarray*}
\small
R_x(\theta) =
\left[\begin{array}{c c}
\cos\frac{\theta}{2} & -i\sin\frac{\theta}{2} \\
-i\sin\frac{\theta}{2} & \cos\frac{\theta}{2}
\end{array}\right]\,\,
R_y(\theta) =
\left[\begin{array}{c c}
\cos\frac{\theta}{2} & -\sin\frac{\theta}{2} \\
\sin\frac{\theta}{2} & \cos\frac{\theta}{2}
\end{array}\right]\\
R_z(\theta) =
\left[\begin{array}{c c}
e^{-i\theta/2} & 0 \\
0 & e^{i\theta/2}
\end{array}\right]\,\,
CNOT =
\left[\begin{array}{c c c c}
1 & 0 & 0 & 0 \\
0 & 1 & 0 & 0 \\
0 & 0 & 0 & 1 \\
0 & 0 & 1 & 0
\end{array}\right]
\end{eqnarray*}

\subsection{Controlled gates}
\label{sec:cgates}

An $n$-qubit circuit processes states represented by $2^n$ amplitudes, $\alpha_y$, with $y \in \{0, 1\}^n$ and $\sum_y|\alpha_y|^2 = 1$. Measuring all qubits of the circuit results in one basis vector with the probability given by the corresponding amplitude, $|\alpha_y|^2$. Quantum gates may act on several qubits simultaneously. A gate operating on $n$ qubits is represented by a $2^n \times 2^n$ complex unitary matrix. One important two-qubit gate is the \emph{controlled-not} CNOT$(c,t)$ gate, where the $c$ qubit conditionally flips the state of the $t$ qubit when set to $\ket{1}$. In general, any quantum gate can be used in a controlled manner, and other versions are controlled-$Z$ (CPHASE), controlled-$V$ (C-$V$) and controlled-$V^\dagger$ (C-$V^\dagger$), where $V^2 = X$.

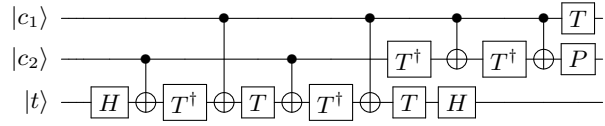
\begin{figure}[t!]
\centerline{
\small{
\Qcircuit @C=.3em @R=.3em {
	\lstick{\ket{c_1}}&\qw&\qw&\qw & \qw & \qw &\qw &\ctrl{2} &\qw&\qw&\qw&\ctrl{2} &\qw&\ctrl{1}&\qw&\ctrl{1}&\gate{T} &\qw\\
	\lstick{\ket{c_2}}&\qw&\qw&\qw& \qw&\ctrl{1}&\qw&\qw&\qw&\ctrl{1}&\qw&\qw&\gate{T^\dagger}&\targ&\gate{T^\dagger}&\targ&\gate{P}&\qw\\
	\lstick{\ket{t}}&\qw&\qw&\qw& \gate{H} & \targ & \gate{T^\dagger} & \targ & \gate{T} & \targ & \gate{T^\dagger} & \targ & \gate{T} & \gate{H} &\qw&\qw&\qw&\qw
	}
}
}
\caption{Toffoli gate using CNOT, $T$, $T^\dagger$ and $H$ gates~\cite[Ch.~4]{NC00}.}
\label{circ:toffoli}
\end{figure}

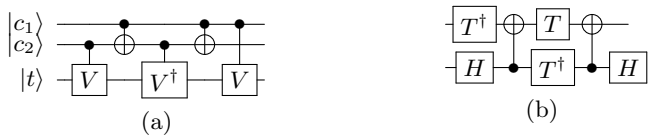
\begin{figure}[t!]
\centering
\subfloat[]{
	\label{circ:toffoli2a}
	\small{
	\Qcircuit @C=.3em @R=.3em {
		\lstick{\ket{c_1}}& \qw & \qw & \ctrl{1} & \qw &\ctrl{1}&\ctrl{2}&\qw\\
		\lstick{\ket{c_2}}& \qw & \ctrl{1} & \targ & \ctrl{1} &\targ&\qw&\qw\\
		\lstick{\ket{t}}& \qw & \gate{V} & \qw & \gate{V^\dagger}&\qw&\gate{V}&\qw
		}
}
}
\hfil
\subfloat[]{
	\label{circ:toffoli2b}
	\small{
	\Qcircuit @C=.3em @R=.3em {
		&\gate{T^\dagger} & \targ & \gate{T} & \targ & \qw\\
		&\gate{H} & \ctrl{-1} & \gate{T^\dagger} & \ctrl{-1}&\gate{H}
		}
}
}
\caption{a) Toffoli gate using CNOT, controlled-$V$ and controlled-$V^\dagger$ gates~\cite[Ch.~4]{NC00}. b) The decomposition of the controlled-V using CNOT, $T$, $T^\dagger$ and $H$ gates}
\end{figure}

Similarly to how arbitrary classical Boolean functions can be constructed entirely from NAND gates, universal quantum computations can be constructed using a discrete set of gates. The universal gate set has to contain at least one coupling operation, and the most often used one is CNOT. A commonly used gate set in fault-tolerant quantum computing is $UGS_{ft}=\{CNOT, H, T\}$~\cite[Ch.~4]{NC00}. There are gate sets that are not universal, an example is the Clifford gate group, generated by the gates $\{CNOT, H, P\}$. Circuits comprised of gates exclusively from 
the Clifford group can be efficiently simulated on a classical computer~\cite{GKtheorem}, but the Clifford group together with the $T$ gate is quantum universal.  The $T$ gate is one of the most expensive quantum gates to implement when QECC and fault-tolerant 
computation is taken into account~\cite{devitt2013requirements,gottesman2013overhead}. Thus, there is ongoing research into reducing the $T$ gate count of synthesised quantum circuits~\cite{jones2012novel,amy2013meet,miller2014mapping}.

\subsection{Reversibility}
The linearity of quantum mechanics has the effect that information can not be erased, therefore, for an arbitrary computation, the number of input qubits equals the number of output qubits. \emph{Reversible circuits}, as presented in~\cite{wille2010towards,saeedi2013synthesis}, are the result of enforcing this requirement on classical Boolean circuits. The interest in classical reversible computing was initially motivated by Landauer's principle, which states that the erasure of information is dissipating energy~\cite{moore2012computing}. The hope was that computers might become more energy-efficient if classical computations would be reversible. Therefore, FANINs and FANOUTs are not allowed into the circuits. The majority of the classical gates are not linear maps. For example the inputs $a$ and $b$ of the $AND(a,b)=c$ gate are impossible to infer from the output $c$. However, the $NOT$ gate is reversible because its output is the negation of the input, and no information is erased.

The reversibility of classical circuits is achieved by the \emph{Toffoli} gate (Fig.~\ref{circ:toffoli}), operating on three bits, where two of them control the bit-flip of the third: \emph{toffoli}$(a,b,c)=(a,b,c \oplus ab)$. Arbitrary classical circuits can be completely constructed using Toffoli gates~\cite[Ch.~3]{NC00}. While a quantum Toffoli performs effectively the same transformation on qubits, the key difference between quantum and reversible circuits is that the Toffoli gate is not universal for quantum computations because universality also require at least the $H$ gate~\cite{aharonov2003simple}. Reversible circuits can be considered restricted quantum circuits operating only on computational basis states. However, it is possible to decompose the Toffoli gate into quantum gates (Fig.~\ref{circ:toffoli} and Fig.~\ref{circ:toffoli2a}). One decomposition (the quantum version) uses the gate set $\{CNOT, H, T\}$ ($T^\dagger=T^7$), while a second decomposition uses the gates $\{CNOT,V,V^\dagger\}$. The second representation will be called \emph{the reversible version} (although the $V$ gate is quantum), because its lower gate cost makes it widely used in the designs of reversible circuits~\cite{saeedi2013synthesis,wille2010towards}, although these costs generally don't account for the true nature of error corrected quantum 
circuits.

\subsection{Information and Gate Teleportation}
\label{sec:igtele}
\begin{figure}[t]
\hfil
\subfloat[]{
	\label{circ:teleportx}
	\Qcircuit @C=1em @R=.7em {
	\lstick{\ket{\psi}} & \ctrl{1} & \measure{X}& \\
	\lstick{\ket{0}} & \targ & \gate{Z}\cwx & \qw&\ket{\psi}
	}
}
\hfil
\subfloat[]{
	\label{circ:teleportz}
	\Qcircuit @C=1em @R=.7em {
		\lstick{\ket{\psi}} & \targ & \measure{Z}& \\
		\lstick{\ket{+}} & \ctrl{-1} & \gate{X}\cwx & \qw&\ket{\psi}
	}
}
\caption{Circuits for teleporting the state of a source qubit to a neighbouring destination qubit.}
\label{circ:teleport2}
\end{figure}
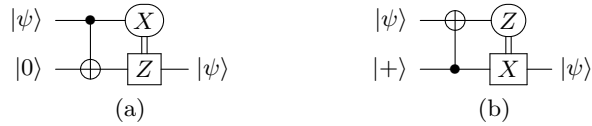

Quantum information (qubit states) cannot be copied~\cite{WZ82}, but there are ways to \emph{move} information from one qubit to another through quantum state \emph{teleportation} (Fig.~\ref{circ:teleport2})~\cite{teleport}. The most general teleportation technique~\cite[Ch.~4]{NC00} is implemented using a slightly different mechanism, but quantum computing models and architectures like~\cite{devitt2009architectural,N14,J13,Y13,FMM13} use the two circuits presented herein.

Each of the circuits requires an ancilla initialised into either $\ket{0}$ or $\ket{+}$. For the first circuit, after applying the CNOT on the states $\ket{\psi}=a\ket{0}+b\ket{1}$ and $\ket{0}$, the two-qubit state will be $a\ket{00}+b\ket{11}$. The measurement of the input qubit, in the $X$-basis
 is probabilistic, and depending on its result the final state of the ancilla will be either $\ket{\psi_1}=a\ket{0}+b\ket{1}$ if $\ket{+}$ is measured, or $\ket{\psi_2}=a\ket{0}-b\ket{1}$ for $\ket{-}$. The execution of the second circuit, where instead a $Z$-basis measurement is used, 
 will result in the state of the ancilla being $\ket{\psi_3}=a\ket{0}+b\ket{1}$ after measuring $\ket{0}$, or $\ket{\psi_4}=a\ket{1}+b\ket{0}$ after measuring $\ket{1}$. For both teleportations the final state is the desired one with $50\%$ probability ($\ket{\psi_1}$ and $\ket{\psi_2}$), while otherwise correctional gates are required, because $\ket{\psi_1}=Z\ket{\psi_2}$ and $\ket{\psi_3}=X\ket{\psi_4}$. The corrections are a direct result of the measurements being probabilistic. The correction mechanism is illustrated in the circuit diagrams by the double vertical lines connecting the measurements to the $X/Z$ gates, indicating a classically controlled gate of either $X$ or $Z$.

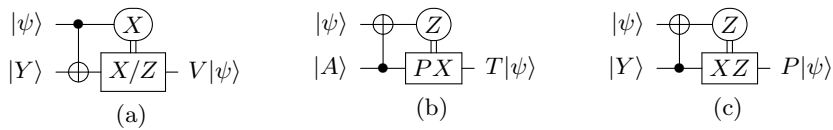
\begin{figure}[t]
\hspace{1.2cm}
\subfloat[] {
	\label{circ:ftv}
	\Qcircuit @C=0.5em @R=.5em {
		\lstick{\ket{\psi}} & \ctrl{1} & \measure{X}\\
		\lstick{\ket{Y}}&\targ& \gate{X/Z} \cwx&\qw&&&V\ket{\psi}
	}
}\hspace{1.7cm}
\subfloat[] {
	\label{circ:ftt}
	\Qcircuit @C=0.5em @R=.5em {
		\lstick{\ket{\psi}} & \targ & \measure{Z}\\
		\lstick{\ket{A}}&\ctrl{-1}& \gate{PX} \cwx&\qw&&&T\ket{\psi}
	}
}\hspace{1.7cm}
\subfloat[] {
	\label{circ:ftp}
	\Qcircuit @C=0.5em @R=.5em {
		\lstick{\ket{\psi}} & \targ & \measure{Z} \\
		\lstick{\ket{Y}}&\ctrl{-1}& \gate{XZ} \cwx&\qw&&&P\ket{\psi}
	}
}
\caption{Teleported rotational gates using the magic states $\ket{A},\ket{Y}$. a) The teleported $V$ gate; b) The teleported $T$ gate; c) The teleported $P$ gate.}
\label{circ:ftcircs}
\end{figure}

Information teleportation is a linear transformation of the destination qubit, such that its state is exactly the state of the source, but quantum gates are linear transformations, too. It follows that it is possible to construct teleported versions for single-qubit quantum gates. Such constructs are commonly used in the fault-tolerant implementation of quantum gates. The teleportation-based gate circuits for the $V$, $T$ and $P$ gates are shown in Fig.~\ref{circ:ftcircs}. The teleportations are again probabilistic and the output state requires corrections (derived in~\cite{paler2014software}). Gate teleportations are based on \emph{magic states}~\cite{bravyi2005universal} like $\ket{Y}=\frac{1}{\sqrt{2}}(\ket{0}+i\ket{1})$ and $\ket{A}=\frac{1}{\sqrt{2}}(\ket{0}+e^{i\frac{\pi}{4}}\ket{1})$. The utilization of magic states and the above teleportation circuits 
is that they can be implemented using fault-tolerant QECC through a process known as state distillation~\cite{bravyi2005universal,FMM13}, which accounts for the majority of resources necessary for a large-scale error corrected algorithm~\cite{devitt2013quantum}.  

The $R_z(\pi/4)^\dagger=T^\dagger = R_z(-\pi/4)$ rotation is implemented using the same circuit as the gate $T$, the only difference being the interpretation of the measurement result in terms of any subsequent correction. Because the $T$,$T^\dagger$,$V$ and $V^\dagger$ gates can be implemented by teleportations, it follows that the Toffoli gate (in both its quantum and reversible versions) can be decomposed into teleportation sub-circuits.

The magic states in the construction of fault-tolerant gates are assumed to be high-fidelity (As high as the fidelity 
of the underlying quantum information protected by the QECC). Otherwise, high-fidelity instances are obtained after \emph{distilling} multiple low-fidelity states using circuits consisting entirely of CNOTs and measurements~\cite{bravyi2005universal}. For example, the distillation of a single 
$\ket{Y}$ state from low-fidelity $\ket{Y}$ ancillae is reported in~\cite{bravyi2005universal}, reducing the infidelity, $p$, of the output from 
$O(p)$, $p<1$, of the seven inputs to $O(p^3)$ on the output.

\section{The ICM Representation}
\label{sec:materials}

In state-of-the-art fault-tolerant quantum circuits, two sources of non-determinism can be distinguished. First, errors can occur during calculation due to undesired interaction with the environment. The errors are handled by quantum error-correcting codes~\cite{devitt2013quantum}. Second, as mentioned above, the realisation of gates by teleportation is inherently probabilistic. The outcome of the gate application is correct with 50\% probability and requires a correction with 50\% probability even in absence of errors.

Circuit gate dynamics, as presented in Section~\ref{sec:intro}, is the consequence of applying specific quantum gates (e.g. $T$) by teleportation. Correctional gates may or may not be required, depending on the outcome of a measurement that is only available when the circuit is being executed. A further source of non-determinism is error-correction, which is not considered herein and is handled at a lower level in the overall design stack of a quantum circuit~\cite{FMM13}. 

A circuit with a dynamic gate list is difficult to execute on a quantum computer, and is furthermore difficult to optimise. This section introduces the ICM representation, which replaces the non-deterministic gate dynamics with an exact gate list. The resulting circuit still contains correctional mechanisms, but these are controlled by measurement results of introduced ancillae and active feedforward determining subsequent measurement choices. We essentially fan-out using extra ancillae to remove the complication of dynamic circuit construction with fault-tolerant and reversible quantum circuits.

\subsection{Non-deterministic Resource Requirements}
\label{sec:nondet}

Gate corrections may or may not be required after each teleportation. They consist in applying $X$, $Y$, $Z$ or $P$ gates to the calculated result. Therefore, the total number of gates in the circuit depends on the number of corrections, and this number is not known \emph{a priori} because the need for corrections is determined only during circuit execution (each individual teleportation has 
a 50:50 chance of each ancilla measurement result, so the possibilities grow exponentially in the number of teleported gates). Moreover, corrections require an introduction of additional ancillae qubits, thus making the computation total number of qubits unpredictable as well.

It can be shown that $X$, $Y$ and $Z$ corrections (Pauli corrections) do not have to be addressed immediately in a quantum way after an unsuccessful gate application. Instead they can be postponed to the end of calculation using \emph{Pauli tracking}~\cite{paler2014software} and 
instead of applying an active quantum gate to the data, we simply reinterpret the meaning of the classical measurement results. However, this technique does not apply to $P$ corrections necessary for implementing the $T$ gate (Section~\ref{sec:igtele}).  This is because the $P$ 
correction does not commute through either the $H$ gate of the target of a $CNOT$ gate in a straightforward manner and changes the 
probability distribution of subsequent $X$-basis measurements. 

For example, in the teleported $T$ gate (Fig.~\ref{circ:ftt}), applying a CNOT on two qubits $\ket{t}=\frac{1}{\sqrt{2}}(\ket{0}+r\ket{1})$ (where $r=e^{\frac{i\cdot\pi}{4}}$) and $\ket{q}=a\ket{0}+b\ket{1}$ results in $\ket{qt}=(a\ket{00}+ar\ket{11}+b\ket{10}+br\ket{01})/\sqrt{2}$. The $\ket{0}$ result of the first qubit's $Z$-measurement will result in the second qubit's state as if it were directly rotated by $T$: $a\ket{0}+br\ket{1}$. If the measurement result is $\ket{1}$, the state is $ar\ket{1} + b\ket{0}$, which after a $PX$ correction is required~\cite{paler2014software}, and it can be applied using the circuit from Fig.~\ref{circ:ftp}.

The $P$ correction requires us to dynamically change the circuit being executed as this correction cannot be classically tracked.  A second ancilla  is introduced in the $\ket{Y} = \ket{0}+i\ket{1}$ state, a CNOT applied between the ancilla and the state to be corrected, and the input is measured according to Fig.~\ref{circ:ftp}. For an $n$-qubit circuit $C$ with a gate list $GL(C)$, each probabilistic $P$ correction increments the number of qubits by one, and inserts a $P$ gate into the gate list.

The problem of applying the $P$ gate dynamically is solved by introducing into the circuit the possibility to operate both a teleported identity gate, used when no correction is needed, and a teleported $P$ gate. Similarly to a classical demultiplexer the measurement result of the teleported $T$ gate is used to decide, at run-time, whether $I$ or $P$ gate is applied. Finally, after performing either the $I$ or $P$ correction, the corresponding state has to be routed to a single qubit. This is realised by classically controlled teleportations in a manner similar to a classical multiplexer with the select signal being the measurement result of the teleported $T$ gate. Classically controlled teleportations were described in~\cite{fowler2012time}, and a circuit using these mechanisms will have a fixed number of qubits and a determined gate list. Compared to a dynamically changing circuit, these are larger, but the predictability of these parameters is useful for circuit optimisation.

\begin{figure}[t]
\small
\centerline{
 \Qcircuit @C=.5em @R=.4em {
		\lstick{\ket{\psi}}& \ctrl{1} & \targ & \qw & \measure{Z} & \measure{X} &&&&&&\lstick{\ket{\psi_1}}&\ctrl{2}&\qw& \measure{X}&\measure{Z}\\
		\lstick{\ket{0}} & \targ & \qw & \targ & \measure{X}&\measure{Z} & &&&&&\lstick{\ket{\psi_2}}&\qw&\ctrl{1}& \measure{Z}&\measure{X}\\
		\lstick{\ket{+}} & \qw & \ctrl{-2} & \qw & \gate{P} &\qw & &&&&&\lstick{\ket{0}}&\targ&\targ&\qw&\qw\\
		\lstick{\ket{+}} & \qw &\qw &\ctrl{-2} & \gate{I} & \qw & &&&&&\\
\\
&&(a)&&&&&&&(b)\gategroup{1}{5}{2}{5}{.4em}{--}\gategroup{1}{6}{2}{6}{.4em}{--}\gategroup{1}{15}{2}{15}{.4em}{--}\gategroup{1}{16}{2}{16}{.4em}{--}
	}
}
\caption{Teleportations: a) Selective destination; b) Selective source~\cite{fowler2012time}.}
\label{circ:selective}
\end{figure}
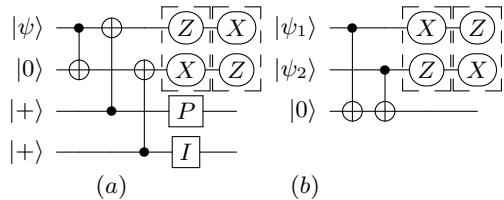

For \emph{selective destination teleportation} (Fig.~\ref{circ:selective}) the first group of measurements ($Z_1X_2$ where the subscripts indicate the qubit's number) will teleport $\ket{\psi}$ on the third qubit where it will be corrected by $P$. The second group of measurements ($X_1Z_2$) will teleport the state to the fourth qubit where the trivial correction $I$ is applied, thus leaving the state unchanged. In the \emph{selective source teleportation} the $X_1Z_2$ measurements will select $\ket{\psi_1}$ for teleportation on the third qubit, while the second measurement group ($Z_1X_2$) will teleport $\ket{\psi_2}$~\cite{fowler2012time}. The selective teleportation circuits require only Pauli corrections, which are not shown in the diagrams, because their application can be postponed to the end of the computation and classically tracked.

As a consequence, Pauli tracking can reduce but not completely eliminate the non-determinism of fault-tolerant circuits. This implies that standard synthesis methods which optimise gate count and/or number of qubits are not applicable to teleportation-based quantum circuits because these numbers are not well-defined. It is possible to circumvent the non-determinism by using ``conditional-identity construction'' which results in the maximal possible number of gates. The initial gate dynamics of a circuit, with all the classically controlled corrections replaced by classically controlled teleportations, is interpreted as the dynamics of the measurements .

\subsection{ICM Correctness and Construction}

The role played by the structured representation of circuits was recognised in~\cite{aaronson2004improved}, where stabiliser circuits were decomposed into a canonical sequence of sub-circuits constructed from a single type of gates. In the context of fault-tolerant quantum computing, the systematic derivations of the circuits~\cite{zhou2000methodology} uses teleportation sub-circuits, too. However, the combination of the fault-tolerant constructions with the regular gate decompositions~\cite{shende2006synthesis}, required for efficient synthesis algorithms, is limited by the realistic requirements of future quantum computing architectures. Nevertheless, structured mapping techniques between various architectures were investigated in~\cite{hirata2011efficient,cheung2007translation}. These approaches were targeted at specific quantum hardware properties, such as nearest-neighbour interaction between qubits, but fault-tolerant constructions were not specifically addressed.

ICM is a structured representation, which consists in the regular representation of arbitrary quantum and reversible circuits using the $UGS_{ft}$ gate set, where the single-qubit rotational gates are teleportation-based. Circuits are transformed into the ICM representation after decomposing all non-$UGS_{ft}$ gates into $UGS_{ft}=\{CNOT, H, T\}$ component gates, and simultaneously introducing, where necessary, selective source and destination teleportation circuits into the resulting circuit.

The correctness of the ICM representation is based on the observation that the teleported gate circuits (Figs.~\ref{circ:ftp},\ref{circ:ftt} and \ref{circ:ftv}) and the selective teleportation circuits (Fig.~\ref{circ:selective}) consist entirely of qubit initialisations, CNOT gates and qubit measurements. Thus, decomposing an arbitrary circuit into elements that can be expressed entirely using the above mentioned sub-circuits, will consist only of initialisations, CNOTs and measurements. The circuit from Fig.~\ref{circ:selective}a can be rewritten, such that the $P$ gate will not be directly applied: in general, $R_z$ rotations (e.g. the $P$ gate) commute with the control of CNOT gates~\cite[Ch.~4]{NC00}. As a result, the $P$ gate can be moved on the left side of the CNOT, and $P\ket{+}=\ket{Y}$. The third qubit from Fig.~\ref{circ:selective}a will be initialised into $\ket{Y}$ instead of $\ket{+}$.

The ICM representation of an arbitrary quantum circuit is the result of applying algorithm presented in this paper. The algorithm is taking a circuit composed of gates from the set $\{$\emph{Toffoli}, $CNOT, C\mbox{-}V$ and $C\mbox{-}V^\dagger, H, P, T\}$, and performs pattern replacements resulting in the circuit $CICM$ (Line 1) consisting of gates from $UGS_{ft}$. The Toffoli gates are decomposed into single qubit rotations (either $\{V,V^\dagger\}$ or $\{H,T,T^\dagger\}$) and CNOT gates. The Hadamard gates are replaced with the series of $Z$- and $X$-axis rotational gates ($P$ and $V$ gates). Afterwards, each $P$ and $V$ gate is replaced using the corresponding teleportation-based gate implementations from Figs.~\ref{circ:ftp},\ref{circ:ftt},\ref{circ:ftv}. The effect of replacing a gate $G$ acting on qubit $i$ is that an ancilla is introduced on the position $i+1$. Thus, all the gates following the initial application of $G$ on $i$ are moved to $i+1$ (Line 20).

The ICM representation is obtained by \emph{moving} all the single-qubit measurements to the end of the circuit, and all the ancillae initialisations to the beginning of the circuit. The middle part of the resulting circuit consists entirely of CNOT gates. The single qubit measurements are then temporally staggered (e.g. Fig.~\ref{fig:papert1}), such that the results of previous measurements determine the basis choices for subsequent measurements to teleport data to pre-prepared ancillae. In the case of the teleported $T$ gate, this procedure dictates to either apply $P$ gate corrections or not, as required.

\subsection{Resource Analysis}

Transforming arbitrary quantum and reversible circuits into the ICM representation requires the introduction of supplemental ancillae, CNOT gates and measurements. The obtained representation is an augmented version of the initial circuit, and there is a constant resource overhead associated with each gate transformation. In the following the gate cost of implementing a sub-circuit (gate) $S$ is represented by $gc(S)$, and the ancilla cost is denoted $ac(S)$.

\subsubsection{Theorem:} The ICM representation of a quantum circuit $C$ with $n_T$ $T$ gates, $n_P$ $P$ gates, $n_V$ $V$ gates, $n_H$ Hadamard gates and $n_{Tf}$ Toffoli gates requires $ac(C)=5n_T + n_P + n_V + 3n_H + 42n_{Tf}$ ancillae and $gc(C)=6N_T + n_P +n_V + 3n_H + 55n_{Tf}$ additional gates.

\subsubsection{Proof:} The central quantum gate is $T$, which requires $ac(T)=5$ ancillae and $gc(T)=6$ CNOTs. One of the ancillae is the one initialised into $\ket{A}$, three other ancillae are used for the selective destination teleportation sub-circuit, and, finally, the fifth ancilla is introduced for the selective source teleportation and represents the output of the teleported $T$ gate.

The $P$ and the $V$ gates introduce a single ancilla $ac(P)=ac(V)=1$ initialised into the $\ket{Y}$ state, and because the teleportation circuits require a single CNOT $gc(P)=gc(V)=1$. The Hadamard gate being implemented as a sequence of $P$ and $V$ gates generates a gate cost of $gc(H)=3gc(P)=3$, and an ancilla cost of $ac(H)=3ac(P)=3$.

The quantum version of the Toffoli gate (denoted \emph{Toffoli}$_q$) decomposition contains $6$ CNOTs, $7$ $T$ gates, one $P$ and two $H$ gates (Fig.~\ref{circ:toffoli}), and thus $gc($\emph{Toffoli}$_q)=6+7gc(T)+(1+2\times 3)gc(P)=55$ and $ac($\emph{Toffoli}$_q)=7ac(T)+(1+2\times 3)ac(P)=42$.

\subsubsection{Note:} The Theorem was formulated for the ICM decomposition of quantum Toffoli gates, but can easily be updated to include the reversible version of these gates (in the following denoted \emph{Toffoli}$_2$). These gates are decomposed into quantum gates, and the initial version contains $2$ CNOT gates and $3$ controlled-$V$ gates (denoted by $CV$), which are further decomposed (Fig.~\ref{circ:toffoli2b}) into 2 Hadamard gates, 3 $T$ and 2 CNOTs. Therefore, because $gc(CV)=2gc(H)+3gc(T)+2=26$ and $ac(CV)=2ac(H)+3ac(T)=21$, the gate cost of the reversible Toffoli is $gc($\emph{Toffoli}$_2)=3gc(CV)+2=80$ and the ancilla cost $ac($\emph{Toffoli}$_2)=3ac(CV)=63$.


\begin{figure}[t!]
\begin{eqnarray*}
\scriptsize
\begin{aligned}
& \textbf{Require: } \text{Circuit } C \text{ composed from } \{Toffoli, CNOT, H, P, T\}\\
&\text{1: }	\text{Circuit }CICM \gets C\\
&\text{2: }	\text{Replace in } CICM \text{ the Toffoli gates with their decomposition (Figure~\ref{circ:toffoli} or Figure~\ref{circ:toffoli2a}})\\
&\text{3: }	\text{Replace in } CICM \text{ the } H \text{ gates with }PVP\\
&\text{4: }	\textbf{forall } P \text{ gates in }CICM\\
&\text{5: }	\quad\text{Introduce the ancilla } a_p \text{ below the qubit having }P\\
&\text{6: }	\quad\text{Construct the circuit for the teleported $P$ gate}\\
&\text{7: }	\quad\text{Move all the gates following the initial } P \text{ onto }a_p\\
&\text{8: }	\textbf{endfor}\\
&\text{9: }	\textbf{forall } V \text{ gates in } CICM\\
&\text{10: }	\quad\text{Introduce the ancilla }a_v \text{ below the qubit having }V\\
&\text{11: }	\quad\text{Construct the circuit for the teleported $V$ gate}\\
&\text{12: }	\quad\text{Move all the gates following the initial }V \text{ onto }a_v\\
&\text{13: }	\textbf{endfor}\\
&\text{14: }	\textbf{forall } T \text{ gates in }CICM\\
&\text{15: }	\quad\text{Introduce the ancilla } a_c \text{ below the qubit having }T\\
&\text{16: }	\quad\text{Construct the circuit for the teleported $T$ gate}\\
&\text{17: }	\quad\text{Introduce }4 \text{ ancillae below the previous ancilla}\\
&\text{18: }	\quad\text{Construct the selective destination circuit where } a_c \text{ corresponds to the first qubit, }\\
&\text{and } s_3 \text{ and } s_4 \text{ are the third and fourth qubits respectively}\\
&\text{19: }	\quad\text{Construct the selective source circuit where } s_3 \text{ corresponds to the first qubit, }\\
& s_4 \text{ to the second qubit, and } a_{out} \text{ is the third qubit}\\
&\text{20: }	\quad\text{Move all the gates following the initial } T \text{ onto the ancilla } a_{out}\\
&\text{21: }	\textbf{endfor}\\
&\text{22: }	\textbf{return } CICM\\
\end{aligned}
\label{alg:1}
\end{eqnarray*}
\end{figure}

State distillation (see Section~\ref{sec:igtele}) is not analysed here, as it is an intrinsic requirement for any type of computation where magic states are required. An exhaustive and complete analysis of the distillation circuits overhead is presented in~\cite{devitt2013requirements} and, as a consequence, the present ICM resource analysis is a continuation of that work.

\section{Discussion}
\label{sec:discussion}

The ICM representation of an arbitrary circuit prepared into a fault-tolerant manner will not affect its properties. Therefore, fault-tolerance statistics will not be discussed. The results of executing the implementation of Algorithm~\ref{alg:1} on circuits from the RevLib benchmark are presented in Table~\ref{tbl:res}. The \emph{EQ} circuits consisted of gates from the set $\{CNOT, C\mbox{-}V, C\mbox{-}V^\dagger\}$ and the \emph{NCT} circuits from the set $\{$\emph{Toffoli}$,CNOT, X\}$. The best-case non-ICM representation consists of the teleportation-based gate construction where no $P$ corrections is required for the $T$ gate. The worst-case non-ICM scenario assumed that all the $T$ gates require the $P$ correction. For other types of gates the corrections can be tracked through the circuit~\cite{paler2014software}, but tracking is not possible for the probabilistic $P$-correction (see Section~\ref{sec:nondet}). In order to illustrate the benefit of the ICM representation the time required for executing the critical path of the decomposed circuits was computed. The model presumed a time cost of $10$ for initialisations, and a cost of $1$ for the CNOTs and the measurements.

It can be seen that the time required by ICM circuits is predictable and better than the worst-case time of circuits before transformation. Note that longer time translates to higher decoherence and more stringent requirements on quantum error-correction.

\begin{table*}[t]
\centering
\caption{Comparison betweeen non-ICM and ICM representation}
\label{tbl:res}
\tiny{
\begin{tabular}{l | c  r r r r r | r r r | r r r | r r r}
\multicolumn{7}{c|}{Original circuit}&\multicolumn{9}{c}{Fault-tolerant circuit}\\
\hline
\multicolumn{7}{c|}{}&\multicolumn{3}{c|}{Best-Case Non-ICM}&\multicolumn{3}{c|}{Worst-Case Non-ICM}&\multicolumn{3}{c}{ICM}\\
\hline
Circuit			&Qub.	&X	&C-X	&Toff.	&C-V	&C-V$^\dagger$	&Ancilla&CNOT	&Time			&Ancilla&CNOT	&Time			&Ancilla&CNOT	&Time\\
\hline
EQ/0410184\_170		&14	&8	&33	&0	&17	&16		&297	&396	&255			&297	&495	&736			&693	&891	&435\\
EQ/3\_17\_15		&3	&1	&3	&0	&2	&4		&54	&69	&56			&54	&87	&177			&126	&159	&100\\
EQ/add16\_175		&49	&0	&32	&0	&48	&16		&576	&736	&287			&576	&928	&1013			&1344	&1696	&551\\
EQ/add32\_185		&97	&0	&64	&0	&96	&32		&1152	&1472	&543			&1152	&1856	&1973			&2688	&3392	&1063\\
EQ/add64\_186		&193	&0	&128	&0	&192	&64		&2304	&2944	&1055			&2304	&3712	&3893			&5376	&6784	&2087\\
EQ/add8\_173		&25	&0	&16	&0	&24	&8		&288	&368	&159			&288	&464	&533			&672	&848	&295\\
EQ/c2\_182		&35	&15	&121	&0	&116	&53		&1521	&1980	&366			&1521	&2487	&1172			&3549	&4515	&661\\
EQ/decod24-v0\_40	&4	&1	&5	&0	&2	&1		&27	&38	&40			&27	&47	&95			&63	&83	&60\\
EQ/decod24-v1\_42	&4	&1	&5	&0	&2	&1		&27	&38	&41			&27	&47	&87			&63	&83	&59\\
EQ/decod24-v2\_44	&4	&1	&5	&0	&1	&2		&27	&38	&41			&27	&47	&95			&63	&83	&60\\
EQ/decod24-v3\_46	&4	&0	&6	&0	&1	&2		&27	&39	&42			&27	&48	&96			&63	&84	&61\\
EQ/fredkin\_5		&3	&0	&4	&0	&1	&2		&27	&37	&41			&27	&46	&94			&63	&82	&59\\
EQ/graycode6\_48	&6	&0	&5	&0	&0	&0		&0	&5	&16			&0	&5	&16			&0	&5	&16\\
EQ/miller\_12		&3	&0	&5	&0	&1	&2		&27	&38	&42			&27	&47	&95			&63	&83	&60\\
EQ/peres\_8		&3	&0	&1	&0	&1	&2		&27	&34	&38			&27	&43	&92			&63	&79	&57\\
EQ/toffoli\_1		&3	&0	&2	&0	&2	&1		&27	&35	&38			&27	&44	&92			&63	&80	&57\\
EQ/toffoli\_double\_3	&4	&0	&4	&0	&2	&1		&27	&37	&38			&27	&46	&94			&63	&82	&59\\
\hline
NCT/0410184\_169	&14	&8	&27	&11	&0	&0		&297	&412	&248			&297	&511	&788			&693	&907	&431\\
NCT/add16\_174		&49	&0	&32	&32	&0	&0		&864	&1152	&440			&864	&1440	&1514			&2016	&2592	&807\\
NCT/add32\_183		&97	&0	&64	&64	&0	&0		&1728	&2304	&840			&1728	&2880	&2938			&4032	&5184	&1559\\
NCT/add64\_184		&193	&0	&128	&128	&0	&0		&3456	&4608	&1640			&3456	&5760	&5786			&8064	&10368	&3063\\
NCT/add8\_172		&25	&0	&16	&16	&0	&0		&432	&576	&240			&432	&720	&802			&1008	&1296	&431\\
NCT/c2\_181		&35	&18	&35	&63	&0	&0		&1701	&2240	&345			&1701	&2807	&1177			&3969	&5075	&631\\
NCT/cnt3-5\_180		&16	&0	&5	&10	&0	&0		&270	&355	&60			&270	&445	&186			&630	&805	&102\\
NCT/graycode6\_47	&6	&0	&5	&0	&0	&0		&0	&5	&16			&0	&5	&16			&0	&5	&16\\
NCT/ham7\_106		&7	&0	&19	&6	&0	&0		&162	&229	&145			&162	&283	&441			&378	&499	&245\\
\end{tabular}
}
\end{table*}

\subsection{Example}

\begin{figure}[t]
\centering
\subfloat[]{
\includegraphics[width=4cm]{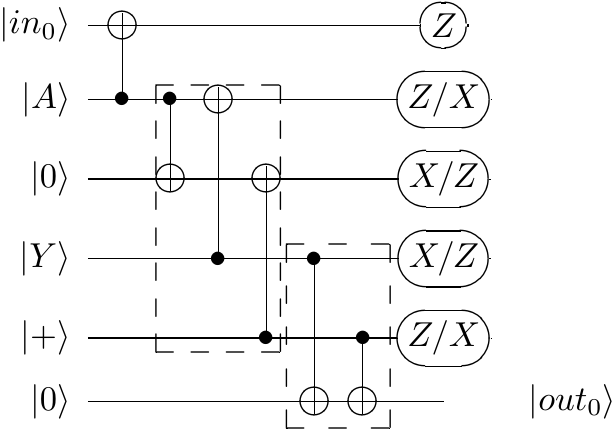}
}
\subfloat[]{
\includegraphics[width=8cm]{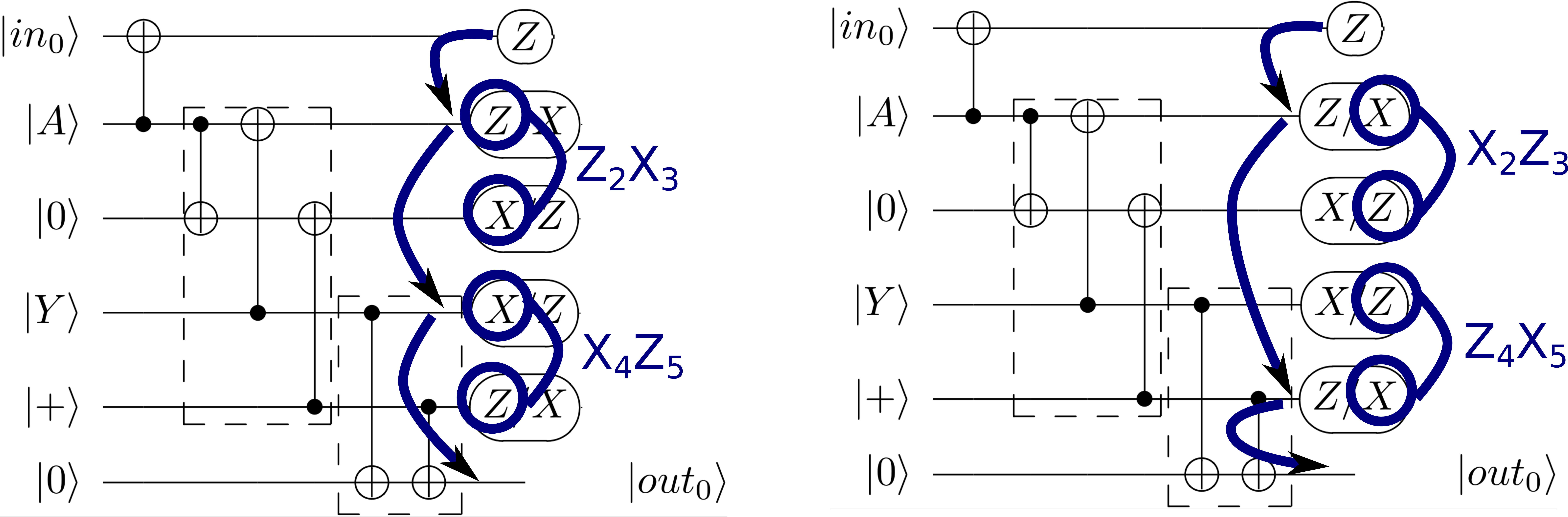}
}
\caption{a) The ICM version of the $T$ gate~\cite{fowler2012time}; b) The arrows sketch the information flow between the ancillas: a $P$ gate correction is the result of the $Z_2X_3X_4Z_5$ measurement pattern, and the $X_2Z_3Z_4X_5$ measurement pattern is used if the teleported $T$ gate application was correct.}
\label{fig:papert1}
\end{figure}

The systematic transformations of the $T$ gate and of the controlled-$V$ gate decomposition from Fig.~\ref{circ:toffoli2b} are presented after applying Algorithm~\ref{alg:1} and obtaining a circuit composed from $UGS_{ft}$ (see Section~\ref{sec:cgates}). The ICM representation of the $T$ gate (Fig.~\ref{fig:papert1}) takes the $\ket{in_0}$ qubit, and after performing the CNOT with the $\ket{A}$ ancilla, selectively teleports (the leftmost group of gates) the intermediary state to either the fourth or the fifth qubit.

The measurement of the first qubit ($Z_1$) is followed by either the measurement pattern $Z_2X_3$ if the result of the teleported $T$ needs a $P$ correction, or the measurement pattern $X_2Z_3$ if the result was correct up to Pauli corrections. The correctness of the teleported gate application is indicated by the measurement result. Applying the $Z_2X_3$ pattern teleports the intermediary state on the output qubit marked by $\ket{out_0}$, and the fourth and fifth qubits are measured using $X_4Z_5$. Otherwise, the measurement $Z_4X_5$ will result in teleporting the state of the fifth qubit on the sixth qubit. The measurement of specific qubit groups depends on the results of previous measurements.

\begin{figure}[t]
\centering
\includegraphics[width=.6\textwidth]{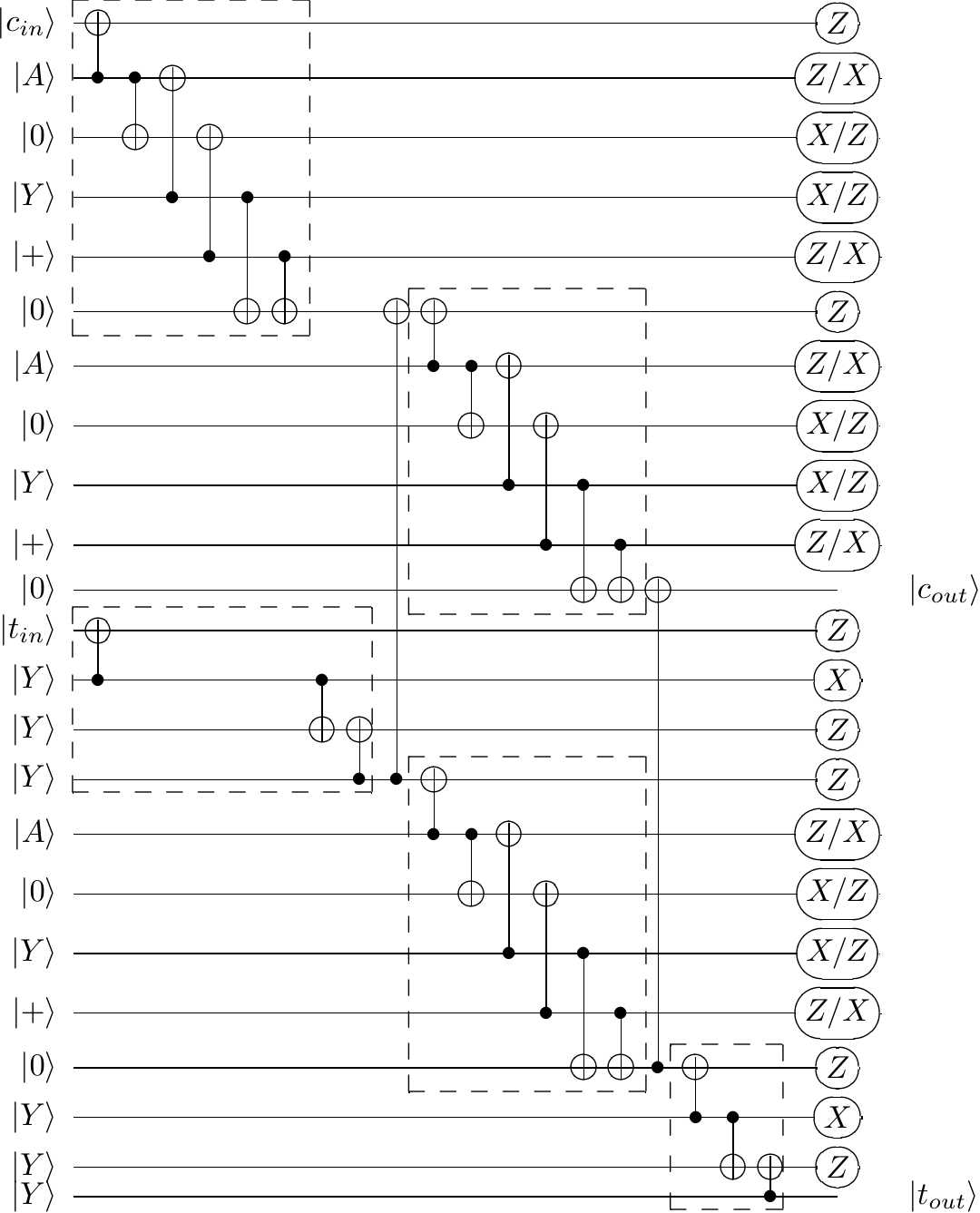}
\caption{The ICM representation of the controlled-$V$ gate. There are three ICM $T$-gate applications (see Fig.~\ref{fig:papert1}) and two ICM Hadamard applications (marked by bounding boxes in which three ancillae are measured using the $ZXZ$ pattern)}.
\label{fig:papert2}
\end{figure}

The controlled-$V$ gate ICM representation (Fig.~\ref{fig:papert2} after applying Algorithm~\ref{alg:1}) has the input states $c_{in}$ (control) and $t_{in}$ (target) and outputs $c_{out}$ and $t_{out}$. The individual decomposition of the single-qubit gates from Fig.~\ref{circ:toffoli2b} is highlighted by the dashed bounding boxes. The boxes containing three CNOTs are implementations of the Hadamard gate where for each constituent sub-gate a CNOT and a $\ket{Y}$-qubit are used. The ancillae introduced by the ICM transformation are affecting the distance between the control and the target of the initial CNOTs (not marked by bounding boxes). The order of the measurements is dictated by the temporal order of the bounding boxes, meaning that the measurements implementing the leftmost $T$ and $H$ can be applied in parallel. Afterwards, the measurements associated to the middle bounding boxes can be again executed in parallel. Finally, the last Hadamard gate from the initial circuit is applied by measuring the last three qubits.

\section{Conclusion}
\label{sec:conclusion}

The usual assumptions made for quantum optimisation techniques do not necessarily hold for the fault-tolerant circuits because of their inherent dynamicity. A regular representation of quantum and reversible circuits was presented starting from the fault-tolerant implementation of quantum circuits. The ICM representation is a consequence of the results presented in~\cite{fowler2012time,paler2014software} and has the potential, when combined with the synthesis method from~\cite{amy2013meet,amy2014polynomial}, to be used for future circuit optimisation techniques.

The results indicate that, while making a quantum circuit fault-tolerant significantly increases its gate count and the number of required ancilla qubits, the ICM representation outperforms direct mapping without enforcing the ICM condition with respect to both predictability and worst-case execution time.  The major advantage of this representation is that it produces a deterministic circuit description for a 
higher level circuit.  A deterministic description is essential to allow for more global circuit optimisations in various error 
corrected implementations.  
Future work will investigate quantum circuit synthesis, optimisation and validation techniques based on the ICM representation.

\section*{Acknowledgements}
SJD acknowledges support from the JSPS Grant-in-aid for Challenging Exploratory Research, NICT, Japan and JSPS KAKENHI Kiban B 25280034.

\bibliographystyle{alpha}
\bibliography{bib}

\end{document}